\documentclass[aps,prd,onecolumn,nofootinbib,longbibliography]{revtex4-2}
\pdfoutput=1
\usepackage[english]{babel}
\usepackage[autostyle]{csquotes}
\usepackage[CJKbookmarks, pdftex, bookmarksnumbered, bookmarksopen, colorlinks, citecolor=cyan, linkcolor=cyan,urlcolor=cyan]{hyperref}
\usepackage{amsmath,amssymb, dsfont,amsthm}
\usepackage{graphicx}
\usepackage{enumitem}
\setlist[enumerate]{topsep=0pt,parsep=-1mm,leftmargin=5mm,}
\MakeOuterQuote{"}
\usepackage{bm}
\usepackage{xspace}

\newtheorem{theorem}{Theorem}
\newtheorem{definition}{Definition}

\renewcommand{\H}{{\mathcal{H}}\xspace}

\newcommand{\ket}[1]{{| #1 \rangle}}
\newcommand{\kket}[1]{{| #1 \rangle\!\rangle}}

\newcommand{\ketbra}[2]{{| #1 \rangle\!\langle #2 |}}

\usepackage{circuitikz}

\newcommand{\discard}{\scalebox{0.5}{$\begin{circuitikz}
\draw node[ground, rotate=180]{};
\end{circuitikz}$}}

\newcommand{\statespace}{\ensuremath{\mathsf{statespace}}\xspace}
\newcommand{\locality}{\ensuremath{\mathsf{locality}}\xspace}
\newcommand{\classicality}{\ensuremath{\mathsf{classicality}}\xspace}
\newcommand{\mediation}{\ensuremath{\mathsf{mediation}}\xspace}

\newcommand{\RR}{{\mathbb{R}}\xspace}

\newcommand{\x}{{x}\xspace}

\renewcommand{\k}{{k}\xspace}

\renewcommand{\S}{{\mathcal S}\xspace}
\newcommand{\T}{{\mathcal T}\xspace}
\newcommand{\E}{{\mathcal E}\xspace}
\renewcommand{\L}{{\mathcal L}\xspace}

\newcommand{\dd}{{\mathrm d}\xspace}
\newcommand{\munu}{{\mu\nu}\xspace}

\newcommand{\bbone}{{\text{\usefont{U}{bbold}{m}{n}\char49}}}

\DeclareMathOperator{\tr}{tr}
\DeclareMathOperator{\diag}{diag}

\begin{document}

\title{Gravity-induced entanglement is not a theory-independent witness of nonclassicality of the gravitational field}

\author{Andrea {Di Biagio}}
\affiliation{Institute for Quantum Optics and Quantum Information (IQOQI) Vienna, Austrian Academy of Sciences, Boltzmanngasse 3, A-1090 Vienna, Austria}
\affiliation{Basic Research Community for Physics e.V., Mariannenstraße 89, Leipzig, Germany}

\begin{abstract}
\noindent Gravity-induced entanglement (GIE) experiments have been proposed as a theory-independent test of the quantum nature of gravity, on the strength of several no-go theorems concluding that GIE detection implies the gravitational interaction is either nonlocal or nonclassical. This is in apparent tension with a growing number of papers claiming that theories with a classical gravitational field do predict GIE. This article resolves the tension by reviewing the assumptions of the no-go theorems and highlighting the information-theoretic, rather than spatiotemporal, nature of the locality assumption. This makes violating locality not particularly problematic. Indeed, quantum electrodynamics and perturbative quantum gravity are nonlocal in the sense of the theorems. Thus, simple detection GIE is not sufficient to rule out a classical gravitational field and we will need quantitative comparison of theoretical predictions in the low-energy quantum gravity regime to test whether the gravitational field is quantum or classical.
\end{abstract}

\maketitle

\noindent Two 2017 papers, one by Bose and collaborators~\cite{bose2017spin} and one by Marletto and Vedral~\cite{marletto2017gravitationallyinduced}, generated a lot of excitement around an old idea by Feynman~\cite{dewitt-morette2011role} of allowing a mass in a superposition of locations to interact gravitationally with another mass to detect gravity-induced entanglement (GIE): the generation of entanglement between two quantum systems as a result of their gravitational interaction.

The excitement around detecting GIE is due to two factors. First, it would be the first direct observation of a quantum-gravity effect: a prediction of perturbative quantum gravity that differs from a prediction of semiclassical%
\footnote{By semiclassical gravity here we mean a theory where the gravitational field is sourced by the expectation value of the energy-momentum tensor. This theory requires non-unitary evolution (spontaneous collapse) to be compatible with observations~\cite{page1981indirect} and, even then, it presents various difficulties with locality and signalling~\cite{gisin1989stochastic,bahrami2014schr}. It is however well-behaved in certain regimes and, indeed, it is the effective framework actually used in essentially all current astrophysical and cosmological applications as well as in precision tests of general relativity~\cite{wallace2021quantum,wallace2022quantum}.} %
gravity~\cite{aleksandergruca2024correlations} testable using near-term technology~\cite{bose2017spin,delic2020cooling,westphal2021measurement,margalit2021realization,aspelmeyer2022avoid,panda2024measuring,bose2025spinbasedpathwaytestingquantum}, making it a concrete target for this generation of experimental physicists. Such a concrete proposal for testing quantum gravity without probing particle collisions at the Planck energy kickstarted a wave of new research into finding other possible low-energy signatures of quantum gravity, sometimes called ``table-top'' quantum gravity~\cite{howl2021nongaussianity,parikh2020noise,toros2024loss,tobar2024detecting,chen2024quantum,NANDI2024138988,higgins2024truly,lantano2025angular,strasser2025evidencing}.

\begin{figure}[h]
  \centering
  \includegraphics[width=0.9\columnwidth]{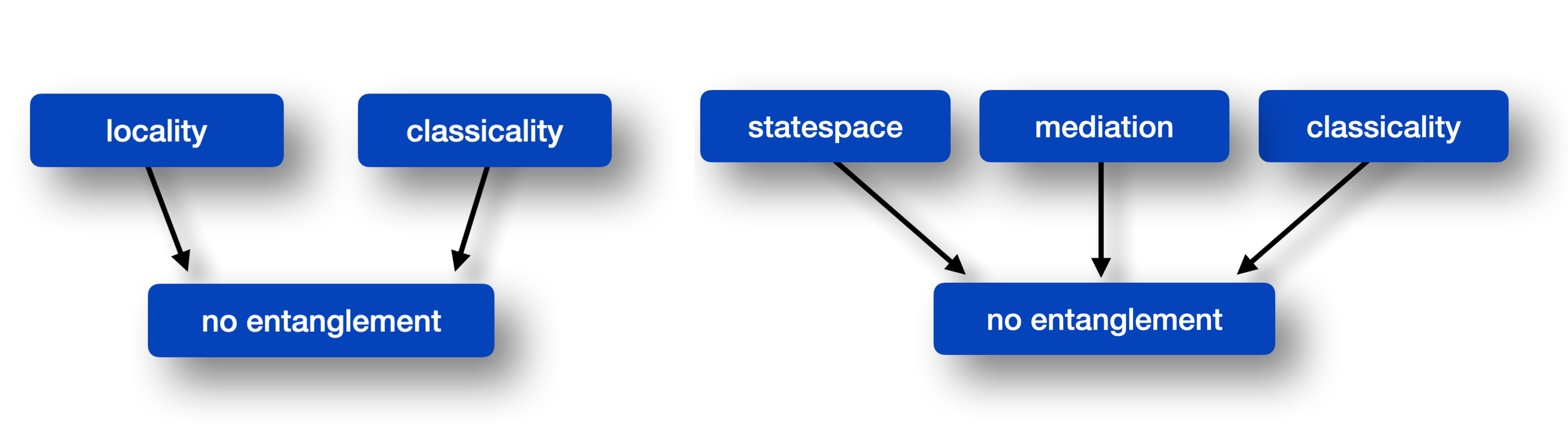}
  \caption{\textit{Two ways to present the LOCC-like no-go theorem assumptions.} \textbf{Left:} The assumptions of the LOCC-like no-go theorems are often presented as \locality and \classicality of the gravitational interaction. However, ``locality'' means different things in different contexts, and this presentation leaves out important nuance of the \locality assumption. \textbf{Right:} A better presentation of the no-go theorems, where the \locality assumption is replaced by two more elementary assumptions: \statespace means that gravity can be assigned an independent statespace like in equation~\eqref{statespace}, \mediation means the evolution of the two masses and gravity factorises as~\eqref{mediation}.}
  \label{fig:assumptions}
\end{figure}

The second reason for excitement, and the main subject of this article, was an argument by Bose \textit{et al.}~and Marletto and Vedral~\cite{bose2017spin,marletto2017gravitationallyinduced}, based on information-theoretic reasoning, that the experiment would provide a theory-independent certification of the nonclassical nature of gravity. This theory-independent argument is based on a generalisation of the well-known quantum-theoretical result that local operations and classical communication (LOCC) channels cannot increase the amount of entanglement between two quantum systems~\cite{horodecki2009quantum}. We will call these generalisations and strengthenings~\cite{bose2017spin,marletto2017gravitationallyinduced,marletto2020witnessing,galley2022nogo,ludescher2025gravitymediated} the \textit{LOCC-like theorems}.\footnote{After the first writing of this article, another theorem appeared~\cite{spalvieri2025local} which makes use of improved assumptions. We will comment on this theorem later.} In the context of low-energy quantum gravity research, these no-go theorems are often stated in words as ``detection of GIE implies that the gravitational interaction is either nonlocal or nonclassical,'' see left side of figure~\ref{fig:assumptions}. Any theory that predicts GIE has to violate either the locality or the classicality assumption of the no-go theorems. A clause is then added: since the principle of locality is unassailable, detecting GIE is sufficient to prove that gravity is nonclassical.

The theorems are correct, and the argument is simple. However, attempts to apply this theory-independent argument to the GIE experiment have been creating controversy ever since it has been proposed~\cite{anastopoulos2018comment,hall2018two,marconato2021vindication,bose2022mechanism,fragkos2023inference,huggett2023quantum,martin-martinez2023what}. Besides conceptual arguments and toy examples, we now know of two concrete theories in which the gravitational field is treated as a classical field and that nevertheless predict GIE. Trillo and Navascu\'es have shown that the Di\'osi--Penrose model---the paradigmatic theory of classical-gravity-induced collapse modification of quantum theory~\cite{diosi1987universal,penrose1996gravity}---predicts GIE~\cite{trillo2025diosipenrose}, while a computation by Aziz and Howl~\cite{aziz2025classical} shows that quantum field theory (QFT) on classical spacetime leads to GIE.\footnote{We must note that there is still an ongoing discussion around the correctness of Aziz and Howl's computation~\cite{diosi2025no,marletto2025classicala,gundhi2026can,vidal2026matterfield}; we will comment on this in Section~\ref{sec:classical}.}

\medskip

Why can classical gravity entangle, despite the no-go theorems? The reason, as we will see, is very simple: the locality assumption of the no-go theorems is not natural in the context of gravity. This assumption is not only violated by quantum-matter-with-classical-gravity theories, but even by relativistic quantum field theory!  Therefore a theory with classical gravity can predict GIE by simply violating this unnatural assumption of locality, which is precisely what happens in the Di\'osi--Penrose theory and in the Aziz--Howl effect.

\pagebreak

\section{Two different notions of locality}

\begin{figure}[h]
  \centering
  \includegraphics[width=0.8\columnwidth]{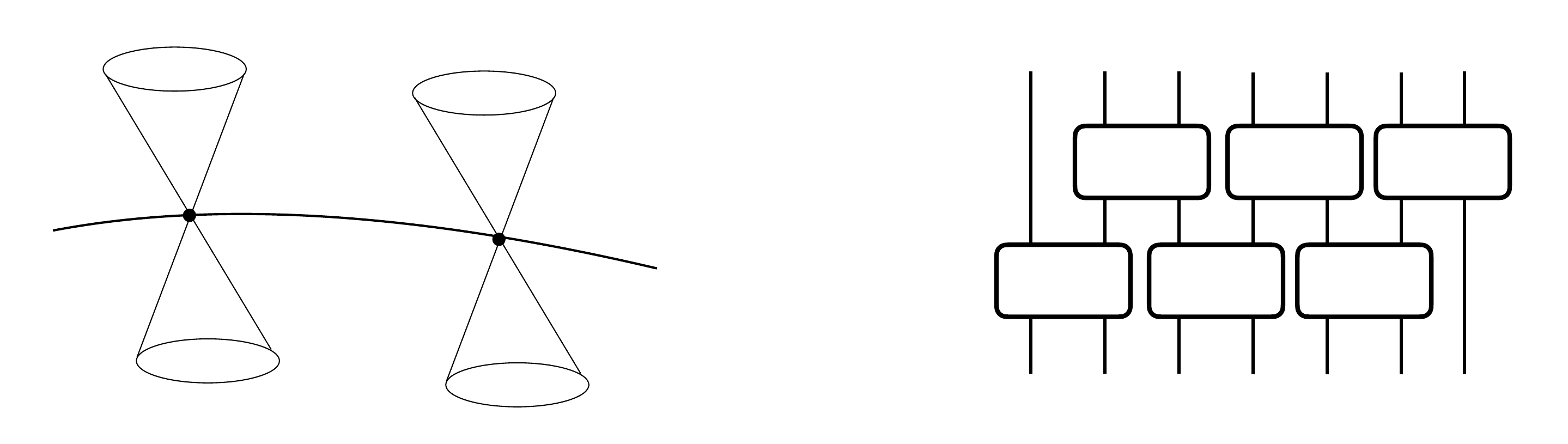}
  \caption{\textit{Two different notions of locality used in physics}. Left, a depiction of a spatiotemporal notion of locality, where information travels within lightcones. Right, a depiction of subsystem, or circuit, locality, where each line represents a system, and the boxes represent operations acting on at most two systems at a time. The first is the notion of locality most salient to relativity and field theory, while the latter is native to information theory and quantum foundations. It is the subsystem notion of locality that is involved in the LOCC-like no-go theorems.}
  \label{fig:locality}
\end{figure}

\noindent Locality is a crucial notion in modern physics. Like time, it has several layers and distinct aspects, captured in various ways by our mathematical formalisations.
When we say "the world is local" we are normally referring to a \textit{spatiotemporal} notion of locality. With the advent of the theories of special and general relativity, the principle of relativistic causality---no signal travels faster than the speed of light---has become a foundational principle of physics. Microcausality---the commutation of operators at spacelike-separated events---is taken as an axiom in quantum field theory, which is arguably the most successful physical theory of all time. Detecting a violation of relativistic causality in the form of faster-than-light signalling would be an epoch-shifting moment, and theories like Bohmian mechanics that only introduce \textit{un}detectable violations of relativistic locality are seen with distrust.

If the no-go theorems showed that any theory of classical gravity that predicted GIE necessarily featured violations of this spacetime notion of locality, they would indeed provide an extremely powerful theory-independent claim. The theorems, however, rely on a \textit{different} notion of locality, based not on spacetime regions, but on \textit{systems}. The interaction between the two particles $A$ and $B$ is local \textit{in the sense of the no-go theorems} if the time evolution of the combined system decomposes in several rounds of interactions between $A$ and $G$ only, and $B$ and $G$ only.  This form of locality is of particular interest in quantum foundations and quantum information research, but it is not well-adapted to field theory.

In the following section, we will analyse the locality assumption of the no-go theorems in detail. We will investigate whether it applies naturally in quantum theory and quantum field theory, and to what extent it can be related to the spacetime notion of locality within quantum theory.

\section{The structure of the no-go theorems}

\begin{table}[h]
\centering
\begin{tabular}{|c|c|c|c|c|}
\hline
\textbf{theorem} & \textbf{framework} & \textbf{postquantum?} & \textbf{no. rounds} & \textbf{dim. systems} \\
\hline
LOCC theorem~\cite{bose2017spin,horodecki2009quantum} & finite-dimensional quantum theory & $\times$ & infinite & finite \\
\hline
Marletto and Vedral (2017)~\cite{marletto2017gravitationallyinduced,marletto2020witnessing} & constructor theory & \checkmark & finite & finite  \\
\hline
Galley \textit{et al.} (2022)~\cite{galley2022nogo} & generalised probabilistic theories & \checkmark & finite & finite \\
\hline
Ludescher \textit{et al.} (2025)~\cite{ludescher2025gravitymediated} & $\mathrm C^*$-algebras & $\times$ & infinite & infinite \\
\hline
\end{tabular}
\caption{\textit{The four main LOCC-like no-go theorems.} The postquantum column denotes whether the metatheoretic framework allows modelling systems that are neither quantum nor classical. All theorems state that a classical system cannot mediate entanglement in a finite number of rounds; however only the two quantum theorems show that the conclusion holds in the limit of infinite number of rounds. Finally, only the recent $\mathrm C^*$-algebra theorem allows the systems to be infinite-dimensional.
}
\label{no-gos_table}
\end{table}

\noindent The four LOCC-like no-go theorems all have the same basic structure: they consider a space of possible theories, then they formulate the two assumptions of \locality and \classicality, and then derive that  two systems $A$ and $B$ initially in a product state cannot get entangled as a result of a local interaction with a classical system $G$. Their difference lies in the space of possible theories and thus the specific implementation of their assumptions. However, they have one thing in common: the \locality assumption is not the spacetime kind of locality introduced above, but a circuit-style notion of locality that, for the purpose of this paper, we will break down into the two notions of $\mathsf{statespace}$ and $\mathsf{mediation}$ so that the no-go theorems can be expressed as in figure~\ref{fig:assumptions}, right.
\begin{theorem}[LOCC-like no-go theorems]
	If two systems $A$ and $B$ initially in a separable state interact via a third system $G$ in a manner that satisfies \statespace, \mediation, and \classicality, then they cannot get entangled as a result of this interaction.
\end{theorem}

\noindent It is outside the scope of this article to review in detail the formalisms and the proofs of these no-go theorems. For our purposes, we will present the no-go theorem in the language of generalised probabilistic theories (GPTs)~\cite{galley2022nogo}, as GPTs are close enough in form to quantum theory so that a reader who has never encountered the framework can just replace every technical element with the more familiar element of quantum theory. We will then briefly comment on the constructor theory~\cite{marletto2017gravitationallyinduced,marletto2020witnessing} and $\mathrm C^*$-algebra~\cite{ludescher2025gravitymediated} theorems.

In both quantum theory and GPTs, a \textit{system} $X$ is represented by a tuple $(\S_X,\T_X,\E_X,\tr_X)$, where $\S_X$ is a convex set of \textit{states},  $\T_X$ is a convex set of \textit{transformations} $\S_X\to\S_X$, and $\E_X$ is a convex set of maps $\S_X\to\RR^+$ called \textit{effects}, together with a distinguished effect $\tr_X\in\E_X$ which represents \textit{discarding} the system and that is used to judge whether a state is normalised or subnormalised. Given two systems $A$ and $B$, there is a recipe to construct the joint system $AB$. For our purposes, this level of generality should be sufficient to present the notions that go into proving the no-go theorems; see~\cite{galley2022nogo,galley2023any,dibiagio2024diagrams} for more details on the use of GPTs to prove no-go theorems on quantum gravity and~\cite{plavala2021general,mueller2021probabilistic} for a more thorough introduction to this powerful framework.  To see the parallel to quantum theory, let $A$ be a system associated with a separable Hilbert space $\H_A$. Then $\S_A$ is the space of density operators on $\H_A$, $\T_A$ is the set of completely positive (CP) trace-nonincreasing maps $\L(\H_A)\to\L(\H_A)$, effects are maps $\rho\mapsto\tr E\rho$ where $E$ is a positive semidefinite operator on $\H_A$, and the distinguished effect is just the trace. A quantum system $A$ can be combined with another quantum system $B$ by repeating the construction with $\H_{AB}=\H_A\otimes\H_B$.

We can now formulate the assumptions of the no-go theorems.
\begin{definition}[\statespace]
	The gravitational interaction between $A$ and $B$ satisfies the \statespace assumption if it can be expressed as an interaction between $A$, $B$, and a third system $G$, where $G$ starts in a separable state and is discarded afterwards.
\end{definition}
\noindent Mathematically, let $T:\S_{AB}\to\S_{AB}$ be the transformation representing the evolution of the systems $A$ and $B$ via the gravitational interaction. This satisfies \statespace if there exists another system $G$ and a map $\tilde T_{ABG}:\S_{ABG}\to \S_{ABG}$ such that, for any state $\rho_{AB}\in\S_{AB}$
	\begin{equation}
  		T_{AB}[\rho_{AB}]=\tr_G \tilde T[\rho_{AB}\otimes\sigma_G],
	\end{equation}
for some $\sigma_G\in \S_G$. We can also represent this assumption via a diagrammatic formula,\footnote{This diagrammatic language to express formulas in linear algebra, initially popularised by Penrose~\cite{penrose2007road}, has the advantage of making evident the compositional structure of maps and it has been proven to be formally equivalent to standard ``one-dimensional'' formulas for a number of branches of mathematics~\cite{coecke2017picturing}.}
\begin{equation}\label{statespace}
  	\vcenter{\hbox{\includegraphics[width=0.3\columnwidth]{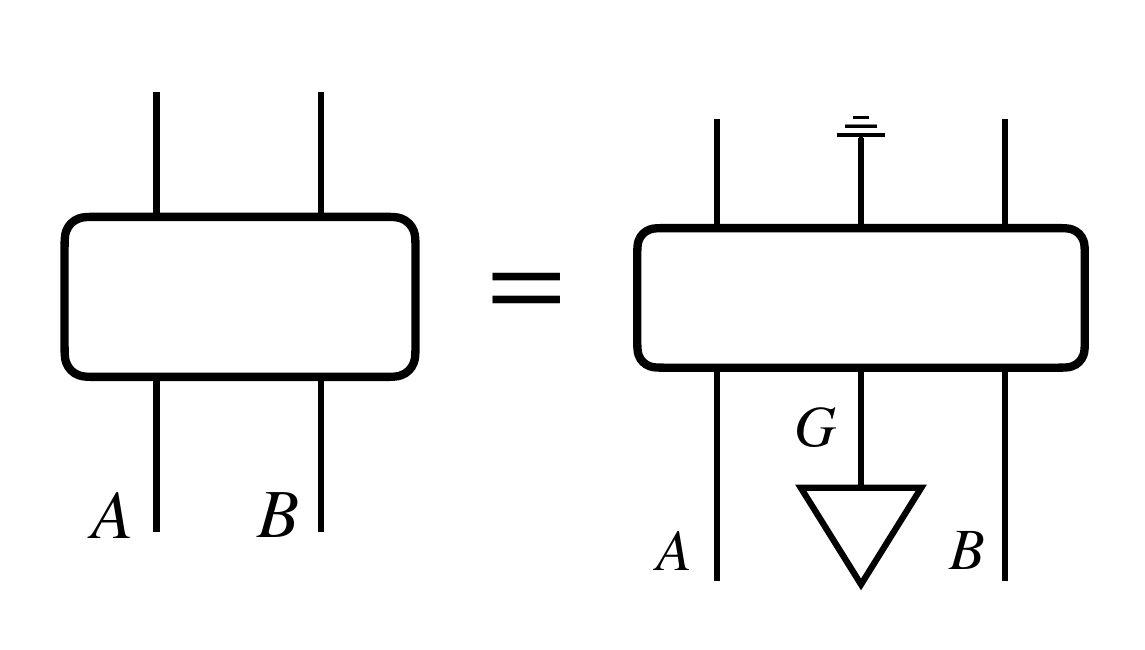}}}\,\,\,.
\end{equation}
In such formulas, each wire represents a system, a box represents a transformation on the systems which flow in and out of the box, a triangle at the bottom of a diagram represents a state for the system, and the ``earth'' symbol $\discard$ represents the discarding operation.

The \mediation assumption is concerned with this map $\tilde T$: it specifies the functional form for the interaction between $A$, $B$, and the gravitational system $G$.
\begin{definition}[\mediation]
	The interaction between $A$ and $B$ satisfies the \mediation assumption if the evolution of the three systems $A$, $B$, and $G$ can be decomposed into several rounds of transformations acting only on $A$ and $G$ or only on $B$ and $G$.
\end{definition}
\noindent Mathematically, this means that there exist sequences $\smash{\{T_{AG}^{(n)}\}}$ and $\smash{\{T_{BG}^{(n)}\}}$ on $\S_{AG}$ and $\S_{BG}$, respectively, such that
\begin{equation}\label{mediation_formula}
  \tilde T_{ABG}=\prod_{n=1}^N\left(\bbone_A\otimes T_{BG}^{(n)}\right)\circ\left(T_{AG}^{(n)}\otimes \bbone_B\right),
\end{equation}
where $\bbone_X$ is used to represent the identity map on $\S_X$. Diagrammatically, this assumption is expressed as
\begin{equation}\label{mediation}
  \vcenter{ \hbox{ \includegraphics[width=0.4\columnwidth]{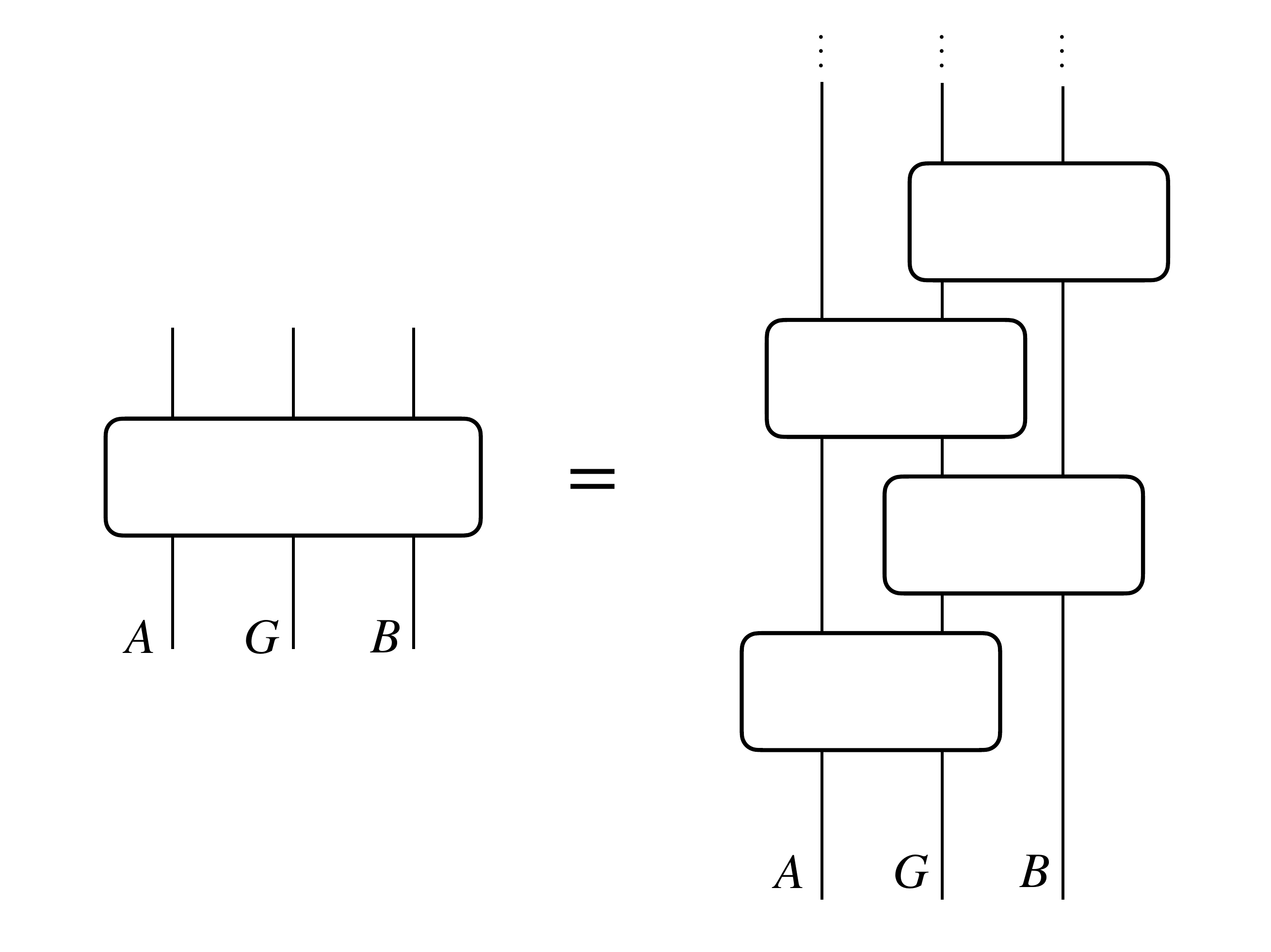}}}\,\,\,,
\end{equation}
where the dots indicate that there could be more than four maps. We note that, while in quantum information theory it is normal to think of time evolution as a finite number of operations in various systems, in mechanics, time evolution is a continuous business. Here it is useful to distinguish between finite-rounds \mediation and infinite-round \mediation. Finite-rounds mediations are the transformations that satisfy~\eqref{mediation_formula} for some finite sets $\smash{\{T_{AG}^{(n)}\}}$ and $\smash{\{T_{BG}^{(n)}\}}$. The infinite-round mediations are the closure of the set of such transformations in some appropriate topology~\cite{horodecki2009quantum,ludescher2025gravitymediated}. 

\begin{definition}
	The gravitational interaction between $A$ and $B$ satisfies \classicality if $G$ is a classical system.
\end{definition}
\noindent The definition of a classical system depends on the details of the framework. A GPT system $G$ is \textit{classical} if its statespace $\S_G$ is a simplex. At first sight this may seem a little unintuitive. However, consider that a \textit{pure state} is a state that cannot be written as a convex combination of other states. Then we see that a classical system is defined as one in which every mixed state can be understood as a statistical mixture in a \textit{unique way}. The analogous definition in quantum theory would be a system whose density matrix $\rho(t)$ is, at all times $t$, diagonal on a \textit{fixed} basis $\ket{i}$, that is $\rho(t) = \sum_i p_i(t)\ketbra ii$.

In the $\mathrm C^*$-algebra framework, a system is associated with a unital $\mathrm C^*$-algebra. There too, one can associate a convex set of states, transformations, and effects, very much like in the GPT framework; a classical system is then associated with a commutative $\mathrm C^*$-algebra. The \locality assumption used in the no-go theorem is completely analogous to~\eqref{mediation}; see \cite[figure 1]{ludescher2025gravitymediated}, although infinite rounds are allowed. The constructor theory framework~\cite{deutsch2015constructor} is a bit different from the previously considered cases as, for example, the sets of states and transformations are not necessarily convex. The notion of locality used in deriving the no-go theorem~\cite[Principle 1]{marletto2020witnessing} is also clearly a notion of locality based on \textit{systems}. The proof of their no-go theorem also assumes \statespace and \mediation\footnote{``By locality, [the generation of entanglement] must be implemented by repeating two elementary steps: first performing a task on $\bold{Q_{A} \oplus M}$ and then on $\bold{M \oplus Q_B}$.''~\cite{marletto2020witnessing}. Here, $\bold{Q_A}$ and $\bold{Q_B}$ are the quantum systems and $\bold M$ is the mediator system responsible for their entanglement.} and a system is classical if it has no complementary observables.

\section{The three limitations of the no-go theorems}

The first remark we need to make is that the theorems do not imply that detection of GIE automatically rules out all theories in which the gravitational field is a classical system. What they rule out are quantum-classical, GPT, and constructor theories in which the gravitational interaction between two systems satisfies \textit{all three} assumptions \statespace, \mediation, and \classicality. Theories that do not easily fit in these frameworks, such as hybrid van Hove theories~\cite{ulbricht2026entanglement} may predict GIE, as could any theory in these frameworks violating any of these three assumptions.

Second, it is evident that the  \locality assumption of the no-go theorems has, a priori, no clear-cut relation to any spatiotemporal notion of locality. In particular, it is not clear at all, at the level of generality at which the no-go theorems are formulated, what is the relation between spacetime locality and the \mediation assumption, not least because there is no notion of \textit{space} or \textit{time} at this level of analysis.  Therefore these theorems should not be understood as pitching the classicality and \textit{spacetime} locality of the gravitational interaction against each other.

Third, a more general point about the application of no-go theorems to make theory-independent (metatheoretic) claims. A powerful no-go theorem presents us with a difficult choice: given certain experimental observations, we are forced to give up at least one cherished assumption. The paradigmatic example is of course Bell's 1964 ``local realism'' theorem~\cite{bell1964einstein,wiseman2015causarum}. There, it was shown that no non-superdeterministic hidden variable model, with no action at a distance, and where measurement reveals the values of the hidden variables, can reproduce the predictions of quantum theory. This puts anyone who is looking for a deterministic hidden-variable completion of quantum theory in a really difficult position, having to accept either superdeterminism or action at a distance. Can the same be said about someone looking for a theory of gravity in which the gravitational field takes on definite values at all points in space and time? This would be the case only if the no-go theorem's \locality assumption is something that is dear at heart to anyone looking for such a theory.

As we will see presently, this is not the case for gravity. On the one hand, in quantum mechanics, Hamiltonians of the form $\hat H_{AG}+\hat H_{BG}$ \textit{do} lead to the infinite-round mediation (the one used in the LOCC and $\mathrm C^*$-algebra theorems) and the finite-round mediation used in the GPT and constructor-theory theorems is found in the relevant regime of the massive scalar field analogue of the GIE experiment. On the other hand, in quantum electrodynamics and linearised quantum gravity, we discover that a theory can be relativistically local while failing to satisfy \mediation, and that whether a theory has \mediation or not is a matter of gauge. Finally, we also point out that because of the gauge constraints, the \statespace assumption itself does not hold in all gauges. Thus, while in some settings the \mediation assumption is fairly natural, this is not the case for a gauge theory like gravity.

\section{Mediation in quantum mechanics and quantum field theory}

\noindent Let us explore to what extent the \mediation assumption holds in our current theories and therefore whether we should expect (or desire) it to hold. We will see that while \mediation can appear quite naturally in quantum theory and scalar quantum field theory, the situation is considerably more complicated in the quantum theory of gauge fields. In particular, we will see that \mediation holds in certain gauges but not in others, which, in turn, means that \mediation cannot be seen as an operational or experimentally-verifiable property of an interaction, at least in quantum field theory. 

\subsection{Quantum mechanics}

\noindent In quantum theory, when we say that two systems $A$ and $B$ are each independently coupled to a third system $G$ we normally mean that the Hamiltonian is written as a sum
\begin{equation}\label{mediation_hamiltonian}
  \hat H = \hat H_{AG} + \hat H_{BG},
\end{equation}
where the first term acts trivially on system $B$ and the second term acts trivially on system $A$ (here, the single-system Hamiltonians have been lumped in either term). Does this coupling imply that the time evolution $\hat U(\Delta t)=e^{-i\hat H \Delta t}$ of this tripartite system is of the mediation form~\eqref{mediation}? That is, does it decompose into a product ${\prod_{n}\left(\hat U_{BG}^{(n)}\hat U_{AG}^{(n)}\right)}$ of unitaries acting nontrivially on $A$ and $G$ only, then $B$ and $G$ only, and so on? Application of the Baker–Campbell–Hausdorff formula shows that this is not generally the case, since
\begin{equation}\label{bhc}
  e^{-i(\hat H_{AG}+\hat H_{BG})\Delta t}= e^{-i\hat H_{BG}\Delta t}e^{-i\hat H_{AG}\Delta t}e^{-\frac12[\hat H_{AG},\hat H_{BG}]\Delta t^2}+{\mathcal O(\Delta t^3)},
\end{equation}
where the commutator $[\hat H_{AG},\hat H_{BG}]$ may very well be an operator that acts nontrivially on all three systems. However, note that we can avail ourselves of the Suzuki-Trotter theorem and write
\begin{equation}
  e^{-i(\hat H_{AG}+\hat H_{BG})\Delta t}= \lim_{N\to\infty} \left(e^{-i\hat H_{BG}\Delta t/N}e^{-i\hat H_{AG}\Delta t/N}\right)^N,
\end{equation}
so that independent couplings as in \eqref{mediation_hamiltonian} do lead to a mediation evolution, but only in the infinite-rounds limit. 

We note that, while the constructor theory and GPT no-go theorems may be generalisable to the infinite-round mediation assumption, at the moment they are only formulated for a finite number of rounds. When do we have a finite number of mediation rounds?

\subsection{Scalar quantum field theory}
\label{sec:scalar}

\begin{figure}[h]
  \centering
  \includegraphics[width=0.75\columnwidth]{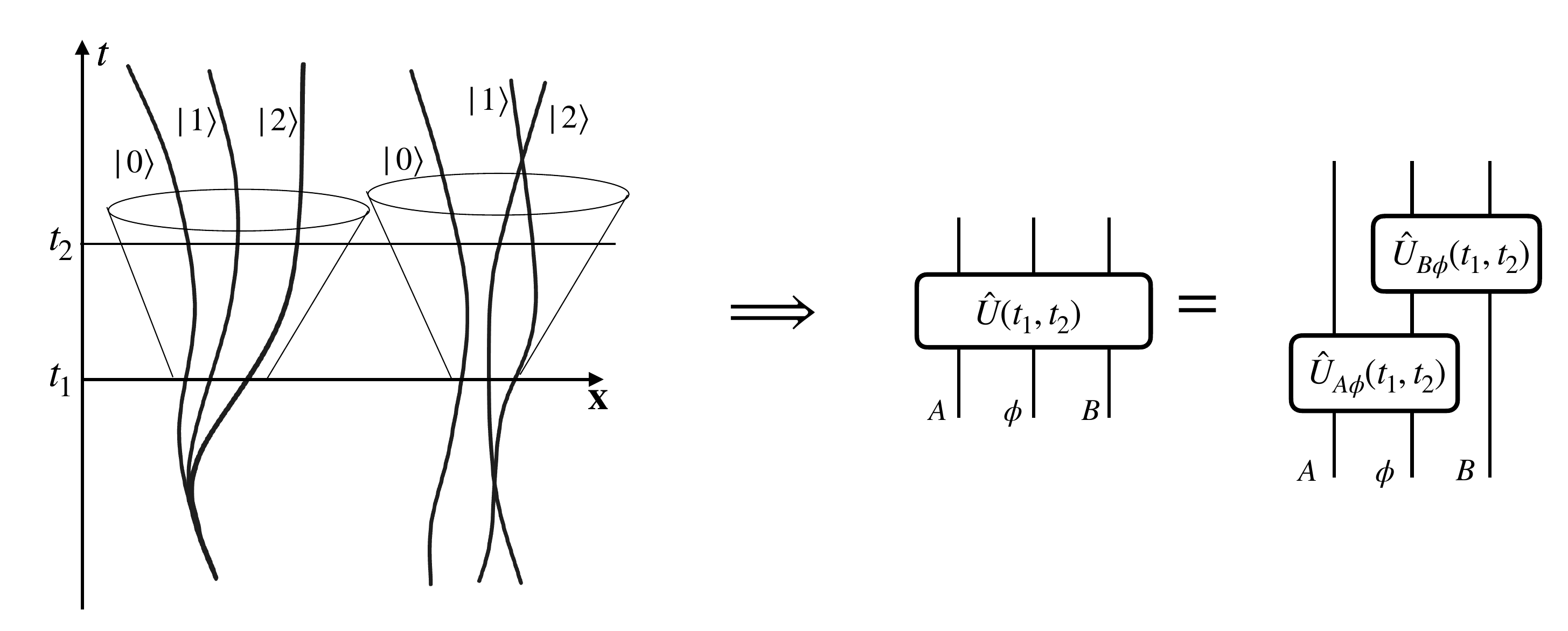}
  \caption{When two particles $A$ and $B$, each coupled to a massive scalar field $\phi$, are placed in a quantum-controlled superposition of localised trajectories, during each time-interval in which all trajectories of one particle are spacelike-separated from all trajectories of the other particle, the time evolution is one round of mediation. Evolving for a finite time interval involves several such rounds.}
  \label{fig:scalar_field}
\end{figure}

\noindent When the middle system $G$ is a quantum scalar field, we can obtain a finite-round mediation evolution. This was shown in detail in~\cite{dibiagio2025circuit} in the case of two Schr\"odinger particles $A$ and $B$ coupled to a real massive quantum scalar field $\phi$. It can be shown that, when the two particles are placed in quantum-controlled superpositions of localised trajectories exactly like in the Bose \textit{et al.}~version of the GIE experiment~\cite{bose2017spin}, the time evolution of the system decomposes into a finite number of mediation rounds, each evolving the system for a finite amount of time; see figure~\ref{fig:scalar_field}.

The main physical ingredient in the result is the relativistic locality of the field, encapsulated by the \textit{microcausality} property: the vanishing of commutators of field operators at spacelike separations,
\begin{equation}
  [\hat \phi(t,\x),\hat \phi(t',\x')]=0~~~\mbox{for $(t,\x)$ and $(t',\x')$ spacelike.}
\end{equation}
Intuitively, this allows the commutators between Hamiltonian densities in the Baker–Campbell–Hausdorff formula~\eqref{bhc} to vanish whenever the time interval $\Delta t$ is such that the particles remain at spacelike separation.\footnote{The story is a little more complicated because a time-dependent Hamiltonian is required to model the quantum-controlled driving which puts and keeps the particles in a superposition of localised states; the evolution is then given by a time-ordered exponential. Nevertheless, the commutation of the field at spacelike separation is a key ingredient leading to the factorisation~\eqref{mediation_scalar}.} More precisely, we obtain that the evolution $\hat U(t_\mathrm{i},t_{\mathrm f})$ between an initial and final time decomposes as $N$ rounds of mediation,
\begin{equation}\label{mediation_scalar}
  \hat U(t_\mathrm{i},t_{\mathrm f}) = \prod_{n=1}^{N}\hat U_{B\phi}^{(n)}\hat U_{A\phi}^{(n)} \hat U_{\phi}^{(n)},
\end{equation}
where each round $n$ evolves the three systems a finite amount of time $(t_{n-1},t_{n})$ and $\smash{\hat U_\phi^{(n)}}$ is the free evolution of the field, while $\smash{\hat U_{A\phi}^{(n)}}$ and $\smash{\hat U_{B\phi}^{(n)}}$ evolve each particle and modify the field to account for the effects that the particle has on the field. In each round these last two operators commute, as one expects from the relativistic locality of the interaction, however $\smash{\hat U_{A\phi}^{(n+1)}}$ will not, in general, commute with $\smash{\hat U_{B\phi}^{(n)}}$ or $\smash{\hat U_\phi^{(n)}}$ and this leads to the build-up of the GIE-like phases.

\subsection{Gauge theories}

\noindent The discussion above seems encouraging: not only quantum theory always provides the infinite-round mediation evolution whenever the systems are coupled to the mediator as in~\eqref{mediation_hamiltonian}, if the mediator is a relativistic scalar field and the particles are localised away from each other then one even obtains finite-rounds mediation. However, gravity is not a simple scalar field theory, it is a \textit{gauge} theory, and this complicates things. As we will see with the example of the electromagnetic (EM) field, the presence of mediation is entirely a matter of gauge, as was already pointed out in~\cite{fragkos2023inference}. 

Let us start by studying quantum electrodynamics (QED) in the radiation gauge, also known as the Coulomb gauge, obtained by gauge-fixing the classical theory in the Coulomb gauge and then quantising. The Hamiltonian of QED in the Coulomb gauge coupled to two charged particles\footnote{We switch to calling the particles $1$ and $2$ instead of $A$ and $B$ in this section as $A$ is normally reserved for the vector potential.} is
\begin{equation}
  \hat H_\mathrm{C}=\hat H_1+\hat H_2+\hat H_{A_\perp}^\mathrm{free}+\hat H_\mathrm{int}+\hat H_{12},
\end{equation}
where the $\hat H_1$ and $\hat H_2$ are the Hamiltonians of the two particles, $\smash{\hat H_{A_\perp}^\mathrm{free}}$ is the free Hamiltonian of the transverse modes and $\smash{\hat H_\mathrm{int}=\int\dd^3x\,\hat{A}_\perp(\x)\cdot (\hat {J}_1(\x)+\hat {J}_2(\x))}$ its local coupling to each particle's current. If this were all there was, we would essentially be in the same situation as in the scalar field case. However, there is also the Coulomb potential term
\begin{equation}
  \hat H_{12}=\iint\dd^3x_1\dd^3x_2\frac{\hat \rho_1(x_1)\hat \rho_2(x_2)}{4\pi|x_1-x_2|},
\end{equation}
which couples the particles directly. This term spoils the mediation form~\cite{anastopoulos2018comment,fragkos2023inference}. We can see this concretely by considering the EM analogue of the GIE experiment. One starts with the two particles localised in some position in the transverse EM vacuum and then slowly manipulates the particles in a superposition of locations. The initial state is ${\ket{\psi_0}=\ket{\mathrm C}_1\ket{\mathrm C}_2\ket{0}_{A_\perp}},$ and by performing the movement arbitrarily slowly, one can make sure that the amplitude of exciting a photon via the current coupling is then arbitrarily low, so that we end up arbitrarily close to the state ${\ket{\psi_1}=\tfrac12\big(\ket{\mathrm L}_1+\ket{\mathrm R}_1\big)\big(\ket{\mathrm L}_2+\ket{\mathrm R}_2\big)\ket{0}_{A_\perp}}$, where $\ket{0}_{A_\perp}$ is the photon vacuum. We then have $\hat H_{A_\perp}^\mathrm{free}\ket{\psi_1}=0=\hat H_\mathrm{int}\ket{\psi_1}$ and
\begin{equation}
  \hat H_\mathrm{C}\ket{\psi_1}=\left(\hat H_1 + \hat H_2 + \frac{q_1q_2}{4\pi|\hat x_1-\hat x_2|}\right)\ket{\psi_1},
\end{equation}
which clearly will not lead to a mediation form, not even in the infinite-round limit. Thus, in Coulomb-gauge QED, two charged particles become entangled in the EM analogue of the GIE experiment, and that entanglement is not mediated.\footnote{Note however that the field has a statespace, namely, the Fock space of the transverse modes. So, \statespace holds but \mediation and \classicality fail.}

From this we learn an important lesson:
\begin{quote}
	\textit{Relativistic causality does not imply mediation.}
\end{quote}
This is because QED in the Coulomb gauge is a perfectly good causal theory that does not feature any superluminal signalling (see~\cite[chapter 9.6]{weinberg2014quantum}, for example) but in which the analogue of the GIE effect happens without a mediator. 

We can however treat the electromagnetic interaction with a different gauge to obtain a mediation evolution. Consider the Gupta-Bleuler quantisation of the EM field~\cite{gupta1950theory,bleuler1950neue,cohen-tannoudji1989photons}, where the Hamiltonian is
\begin{equation}
  \hat H_\mathrm{GB}=\hat H_1+\hat H_2+\hat H_{A}^\mathrm{free}+\hat H_\mathrm{int},
\end{equation}
where $\hat H_{A}^\mathrm{free}$ is the sum of the free Hamiltonians for the four components  $\hat A^\mu$ of the vector potential and the interaction term $\smash{{\hat H_\mathrm{int}=\int\dd^3x\,\hat{A}_\mu(\x)(\hat {J}^\mu_1(\x)+\hat {J}^\mu_2(\x))}}$ locally couples each of these components independently to each of the charged particles. The commutators of the vector potential satisfy $\smash{{[\hat A_\mu(t,\x),\hat A_\nu(t',\x')]=i\eta_\munu D(t-t',x-x')}}$ where $\eta_\munu=\diag(-1,1,1,1)$ and $D(t-t',x-x')$ is the Pauli-Jordan function, which vanishes at spacelike separation.  This means that the situation is very much like the scalar field case in section~\ref{sec:scalar}: the evolution consists of a finite number of rounds of mediation by a quantum system, and thus \mediation holds.

However, note that it is not only the \classicality assumption that fails: the \statespace assumption also fails. This is because in the Gupta-Bleuler formalism we need to impose the Lorenz gauge-condition $(\partial_\mu \hat A^\mu)^{(+)}\kket{\psi}=0$, where $\kket\psi$ denotes a \textit{physical} state, as opposed to a generic vector of the kinematical statespace. Together with the covariant Maxwell equations, this implies Gauss' law on physical states
\begin{equation}
  (\nabla\cdot\hat E-\hat\rho)\kket\psi=0.
\end{equation}
This means that a generic physical state will be entangled with respect to the tensor product structure $\H_1\otimes\H_2\otimes\H_{A}$ of the kinematical statespace.

We focussed on electromagnetism because of its relative simplicity, but the situation is entirely analogous in linearised quantum gravity. In a reduced (or canonical) quantisation of linearised theory, obtained by solving the Hamiltonian constraints before quantisation, the gravitational interaction between matter sources appears through an instantaneous Newton potential~\cite{anastopoulos2018comment}. In this formulation the Hamiltonian therefore contains a direct-interaction term between the matter particles, analogous to the Coulomb term in QED in the Coulomb gauge, and the resulting entanglement is not mediated. One \textit{can} however quantise the linearised metric perturbation $\hat h_{\mu\nu}$ covariantly, in the Gupta-Bleuler formalism based on the de Donder (harmonic) gauge condition $\smash{\partial_\mu h^{\mu\nu}-\tfrac12\partial^\nu h=0}$. In this description, the interaction Hamiltonian 
$\smash{\hat H_{\rm int}\propto\int d^3x\, \hat h_{\mu\nu}(x)\hat T^{\mu\nu}(x)}$ is local in space
and the commutators of the field operators $\smash{\hat h_\munu}$ again vanish at spacelike separation. At the level of the kinematical field variables the evolution therefore decomposes into mediation rounds, exactly as in the scalar-field and Gupta-Bleuler electromagnetic cases. However, as in electromagnetism, the gauge condition together with the linearised Einstein equations implies the gravitational constraints (the analogues of Gauss' law), which relate the metric perturbation to the matter stress-energy tensor. Physical states must therefore satisfy these constraints, and the physical statespace does not factorise with respect to the kinematical tensor-product structure between matter and gravitational degrees of freedom. As a result, generic physical states are entangled with respect to that tensor-product structure and \statespace fails. The difficulties in treating the two masses and the field as independent subsystems are discussed in detail from an algebraic quantum perspective in~\cite{boulle2025subsystems}.

In both electromagnetism and gravity, we see that there is a gauge in 
which \mediation holds but \statespace fails and another gauge where the \mediation assumption fails but \statespace holds. Thus we learn another important lesson:
\begin{quote}
	\textit{The assumptions of the LOCC-like no-go theorems are not gauge-independent.}
\end{quote}
Even assuming linearised quantum gravity, whether the gravitational field is \textit{mediating} the generation of entanglement between two masses or merely \textit{inducing} it depends on the choice of gauge to compute the evolution. This, in particular, means that one cannot operationally enforce \mediation.\footnote{Compare this with spacelike separation during a Bell experiment, which is an operational notion that can, in principle, be verified experimentally by looking at rulers and clocks.}

The more recent theorem by Spalvieri~\textit{et al.}~\cite{spalvieri2025local}, formulated within the framework of lattice gauge models, does not make use of the \statespace assumption. Instead, they formulate a per-sector \mediation assumption that does not assume that the statespaces of the mediator and matter factorise.  While constraints spoil the kinematical Hilbert space factorisation, one can still decompose the physical Hilbert space in a \textit{direct sum} of tensor product Hilbert spaces. The authors then show that a per-sector mediation by a classical system cannot lead to entanglement. This clearly generalises the no-go theorems. However the per-sector \mediation is still gauge-dependent, as it would not hold in the Coulomb gauge.

Let us summarise the main takeaways from this section. The \statespace and \mediation assumptions are gauge-dependent: the field mediates the generation of entanglement in one gauge but not in the other. Relativistic causality and the retarded propagation of signals hold in both gauges, meaning that the \locality assumption of the no-go theorems is independent of relativistic causality. Together, these imply a) that performing a GIE experiment in regimes where retardation is measurable does not enforce \mediation and thus still does not allow us to use the LOCC-like no-go theorems to infer the nonclassicality of the gravitational field, and b) that whether gravity \textit{mediates} the generation of entanglement is not a fact that an experiment can establish, not in a theory-independent way and sometimes not even within a specific theory.

\section{The failure of \mediation and \statespace in theories with classical gravity}
\label{sec:classical}

Let us consider two concrete examples of theories with classical gravity and quantum matter that nevertheless predict gravity-induced entanglement. None of these contradict the LOCC-like no-go theorems: they only invalidate the claim that detection of GIE is incontrovertible proof of nonclassicality of the gravitational field.

Trillo and Navascu\'es have shown that the Di\'osi--Penrose (DP) model~\cite{diosi1987universal,penrose1996gravity} predicts GIE~\cite{trillo2025diosipenrose}. The time evolution of the DP model can be understood as a continuous measurement of the matter distribution, the outcome of which is used to define the value of the classical gravitational field~\cite{tilloy2016sourcing}. This dynamics does not satisfy the \statespace and \mediation assumptions in the Newtonian limit, because the field values are entirely dependent on the outcome of the matter measurements. Even if one would allow for some independence for the gravitational field states, the evolution would still not satisfy \mediation. Either way, the gravitational field in the DP model is classical in the precise sense that it has a well-defined value at all spacetime events. 

The Aziz--Howl (AH) effect is the creation of entanglement between two local excitations of a complex scalar field in the presence of a gravitational field. Each `particle' of the GIE experiment is treated as an $N$-particle excitation of a specific mode of a scalar field,
\begin{equation}
  \ket{\psi_1} = \frac1{2N!}\left((\hat a_{A\mathrm L}^\dagger)^N+(\hat a_{A\mathrm R}^\dagger)^N\right)\left((\hat a_{B\mathrm L}^\dagger)^N+(\hat a_{B\mathrm R}^\dagger)^N\right)\ket0,
\end{equation}
where $\hat a_{A\mathrm L}=(2\pi)^{-3}\int\psi_{A\mathrm L}(k)\hat a_k\,\mathrm d^3k$ annihilates a particle with wavefunction $\psi_{A\mathrm L}$.
This state then evolves according to a Hamiltonian $\hat H=\hat H_0+\hat H_\mathrm{int}$, where $\hat H_0$ is the flat-space free Hamiltonian for a complex scalar field and
\begin{equation}
  \hat H_\mathrm{int}=-\frac12\int\dd^3x\,h_\munu(x)\hat T^\munu(x),
\end{equation}
where $h_\munu$ is the classical perturbation of the metric. Aziz and Howl showed that when $h_{\munu}$ is inhomogeneous, this evolution results in entanglement between the particles, that is, between the two $A$ modes and the two $B$ modes. In their perturbative computation, this is due to a fourth-order Feynman diagram which can be interpreted as the exchange of two virtual matter particles. The effect is due to gravity in the precise sense that this Feynman diagram has no amplitude and no entanglement is generated if $h_\munu$ is constant and homogeneous.

The effect can be understood in non-perturbative terms as the gravitational field activating an interaction between the two excitations and the rest of the modes of the complex scalar field. Indeed, in the non-relativistic limit where $h_\munu(x)=-2\Phi(\x)\,\delta_\munu$, $\hat H_\mathrm{int}$ reads~\cite[Supplementary information eq.~(41)]{aziz2025classical}
\begin{equation}\label{h_int_si}
  \hat H_\mathrm{int}=4\int\dd^3x\,\Phi(\x)\left(\hat\pi(\x)\hat\pi^\dagger(\x)-\frac{m^2}{2}\,\hat\phi^\dagger(\x)\hat\phi(\x)\right),
\end{equation}
where
  $\hat\phi(\x)=\int\frac{\dd^3\k}{(2\pi)^3}\frac{1}{\sqrt{2\omega_\k}}(\hat a_{\k}e^{i\k\cdot\x}+\hat b^\dagger_{\k}e^{-i\k\cdot\x})$
is the full matter field, which contains the modes of particle $A$, particle $B$, and all the other modes, $\hat\pi=\partial_0\hat\phi^\dagger$ is the momentum conjugate to the field, $\hat b^\dagger_\k$ creates antiparticles, and $\omega_\k=\sqrt{\k^2+m^2}$. Expanding~\eqref{h_int_si}, one sees that an inhomogeneous gravitational potential $\Phi$ couples the $A$ and $B$ modes to every other mode of the field and, since these are quantum, \textit{they} can mediate the entanglement, diagrammatically,
\begin{equation}
\vcenter{ \hbox{ \includegraphics[width=0.3\columnwidth]{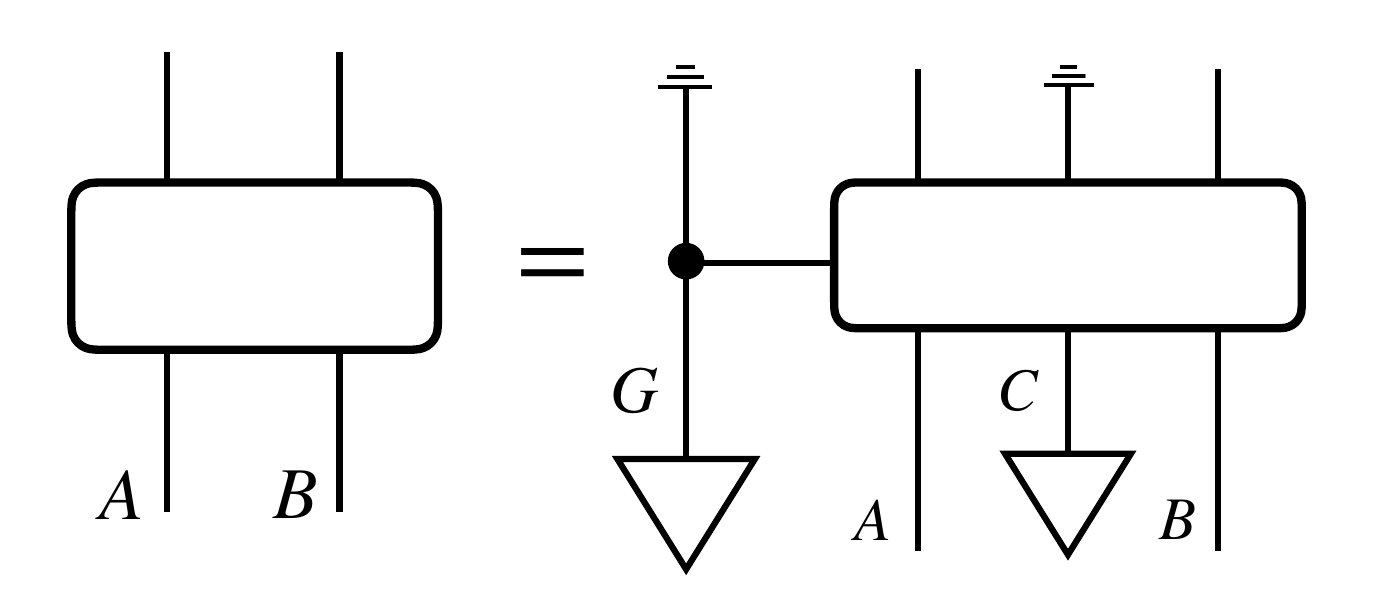}}}\,,
\end{equation}
where we denoted by $C$ the system composed of the rest of the modes of the scalar field besides the $A$ and $B$ ones. In their model, the gravitational field could be either a purely independent background field, or sourced by the expectation value of the energy-momentum tensor~\cite[Supplementary Information Sections 2.1 and 2.2]{aziz2025classical}, or in some other way. So the Aziz--Howl effect is also due to a failure of the \mediation assumption. The entanglement is thus gravity-\textit{induced} but matter-\textit{mediated}: the classical gravitational field activates an effective interaction among the matter modes, and the system that mediates is quantum matter, not gravity.

We should note that the correctness of the Aziz--Howl effect is still under debate~\cite{diosi2025no,marletto2025classicala,gundhi2026can,vidal2026matterfield}, with some papers claiming there must be a mistake in the computation. Soon after the paper came out, two short nonperturbative analyses appeared, arguing that the effect must be an artefact of Aziz and Howl's lengthy perturbative computation~\cite{diosi2025no,marletto2025classicala}. Marletto \textit{et al.}~\cite{marletto2025classicala} argue that given that the $A$ and $B$ wavepackets are nonoverlapping and the fields couple locally at a point, the Hamiltonian splits into two commuting local terms and so cannot entangle. Di\'osi~\cite{diosi2025no} argues instead that the initial factorisation persists in time, appealing to the linearity of the Heisenberg-picture evolution of the mode operators. Both arguments are fundamentally flawed, however, as they overlook that the interaction Hamiltonian is quadratic in the \textit{full} matter field, so the classical potential couples the $A$ and $B$ to the other modes of the scalar field, which can act as mediators, as explained above. A more recent work by Gundhi \textit{et al.}~argues that the entangling term is an artefact of discarded transition amplitudes~\cite{gundhi2026can}. Tibau Vidal and Iyer refuted this analysis: the substitution of Wick contractions by Gundhi \textit{et al.}~is missing a term that, once taken into account, exactly suppresses the off-diagonal terms that would make the phase separable, so that Aziz and Howl are not missing any fourth-order term~\cite[Section IV]{vidal2026matterfield}. As it stands, there is no concrete evidence that the computation is wrong.

One critique often raised against both the DP model and the Aziz--Howl effect concerns how much entanglement they predict in concrete experiments.\footnote{This happens mostly in conversation, but see also~\cite{tang2025mattermediated,xue2026aziz}, for example.} This brings us to the final section of this paper.

\section{Why the GIE experiments are still crucial}

 \begin{table}[h]
\centering
\begin{tabular}{|c|c|c|c|}
\hline
\textbf{theory} & \textbf{GIE?} & \textbf{assumption dropped} & \textbf{good candidate?} \\
\hline
Newtonian QM & \checkmark & \statespace & $\times$ (no gravitational waves) \\
\hline
Schr\"odinger-Newton \cite{bahrami2014schr} & $\times$ & \statespace & $\times$ (possibly inconsistent) \\
\hline
semiclassical gravity and QFT \cite{aziz2025classical} & \checkmark & \mediation & $\times$ (possibly inconsistent) \\
\hline
perturbative QG in Lorenz gauge & \checkmark & \classicality, \statespace & \checkmark \\
\hline
perturbative QG in Coulomb gauge & \checkmark & \classicality, \mediation & \checkmark \\
\hline
mongrel gravity (Oppenheim) \cite{oppenheim2023postquantum,oppenheim2025covariant} & $\times$ & none (see caption) & ? (requires relativistic extension) \\
\hline
Di\'osi--Penrose model~\cite{diosi1987universal,trillo2025diosipenrose} & \checkmark & \statespace, \mediation & ? (requires relativistic extension) \\
\hline
\end{tabular}
\caption{\textit{How different concrete theories sit with respect to the LOCC-like no-go theorems.} ``GIE?'' indicates whether the theory predicts gravity-induced entanglement between two nearby masses in a superposition of locations, ``assumption dropped'' indicates which assumption of the no-go theorems is not satisfied, ``good candidate?'' is a rough assessment of whether the theory has been ruled out either by observation or in some other way. No theory that predicts GIE violates \textit{only} the \classicality assumption. Oppenheim's mongrel gravity drops none of the assumptions, although whether it satisfies \mediation exactly or only approximately depends on the choice of noise kernel~\cite{tilloy2024general}. Here QM stands for quantum mechanics, QFT for quantum field theory, and QG for quantum gravity. }
\label{theories_table}
\end{table}

\noindent  The Bose \emph{et al.}~and Marletto and Vedral papers promised us a simple way to prove once and for all that gravity is nonclassical. Under closer scrutiny, their argument does not go through. In table~\ref{theories_table}, we list a number of theories that have been shown to either predict or not predict the GIE. Besides the arguments in this paper, we can see that the only good theory that would be ruled out by simply \textit{detecting} GIE is Oppenheim's so-called ``mongrel'' or ``postquantum gravity'' theory~\cite{oppenheim2023postquantum,oppenheim2025covariant}, although we should note that, technically, whether Oppenheim's theory satisfies \mediation exactly or only approximately depends on the choice of noise kernel~\cite{tilloy2024general}. We also note that of the theories that predict GIE, none exclusively violates the \classicality assumption.

Regardless of no-go theorems, GIE experiments, carefully analysed and performed, will be a landmark moment in the history of physics because they will allow testing the predictions of various different theories of the gravitational interaction. The theory-independent approach of ruling out entire classes of (existing or hypothetical) theories with a single experiment is a relatively new addition to scientific practice: we have been ruling out theories way before inventing no-go theorems. What we need to do is the good-old empirical testing of theories: carefully compute detailed predictions of concrete theories in specific experimental setups and compare with empirical data.
The alternative theories of classical gravity coupled to quantum matter differ in their quantitative predictions in the GIE experiment regarding how entanglement rates, maximal entanglement, and decoherence rates depend on the various parameters of the experiments, such as the masses of the objects in superposition, their distance, and the duration of the experiment.

How different are the predictions? Consider the regime in which the branch separation $\Delta x$ is small compared to the distance $d$ between the objects. In linearised quantum gravity the negativity\footnote{The \textit{negativity} of a bipartite state $\rho$ is the sum of the absolute values of the negative eigenvalues of its partial transpose~\cite{horodecki2009quantum}. For the effectively two-qubit states relevant here, $\mathcal N=0$ if and only if the state is separable, and $\mathcal N=\tfrac12$ for maximally entangled states.} oscillates between zero and maximal; at early times it grows as $\mathcal N\propto GM^2t\,\Delta x^2/\hbar d^3$, where $M$ is the mass of the objects~\cite{christodoulou2023locally}. Compare that to the Aziz--Howl effect, where $\mathcal N\propto (G^2m^2M^3Rt/\hbar^3 d)^2\Delta x^2/d^2$, with $m$ the mass of the field quanta (the atoms) and $R$ the radius of the objects~\cite{aziz2025classical}. In the Di\'osi--Penrose model, entanglement is generated only for separations below a critical distance $d_c\approx0.85\sigma$ set by the model's free parameter $\sigma$; the negativity then rises to a plateau and slowly decays~\cite{trillo2025diosipenrose}. These theories predict clearly different quantitative and qualitative behaviours.

In the field of low-energy quantum gravity experiments, the LOCC-type no-go theorems seemed to give GIE a special role: while other experiments had to rely on assumptions about different theories, GIE would provide a theory-independent certification of nonclassicality of the gravitational interaction. Now that we have seen that the theory-independent claim does not hold, we can and must continue the search for more effects, such as non-Gaussianity~\cite{howl2021nongaussianity}, graviton noise~\cite{parikh2020noise,toros2024loss} and graviton detection~\cite{tobar2024detecting}, effects due to the non-commutativity of the field~\cite{chen2024quantum}, entanglement from rotational  superposition~\cite{higgins2024truly,lantano2025angular}, and thermodynamical observables~\cite{strasser2025evidencing}. Thanks to the steady increase in experimental control over quantum matter, these effects will allow us to test perturbative quantum gravity against concrete rival theories where gravity is classical. No need for no-go theorems.

 \begin{acknowledgments}
 \noindent I am grateful to Marios Christodoulou, Thomas Galley, \v Caslav Brukner, Carlo Rovelli, Markus Aspelmeyer, and Richard Howl for many discussions about GIE and the LOCC-like no-go theorems. I also wish to thank the organisers and participants of the workshop ``A look at the interface between gravity and quantum theory 2025'' as well as Anne-Catherine de la Hamette for useful and encouraging discussions, and Renato Renner for many helpful questions. Finally, I wish to thank an anonymous referee for suggestions to improve the paper.
 
 	This project was funded within the QuantERA II Programme that has received funding
from the European Union’s Horizon 2020 research and innovation programme under Grant
Agreement No 101017733, and from the Austrian Science Fund (FWF), projects I-6004 and
ESP2889224 as well as Grant No. I 5384.
 \end{acknowledgments}

\bibliography{refs.bib}
\end{document}